\documentclass{WileyMSP-template}

\begin{document}


\title{Image transmission through a dynamically perturbed multimode fiber by deep learning}

\maketitle


\author{Shachar Resisi,}
\author{Sebastien M. Popoff,}
\author{Yaron Bromberg*}

\dedication{}

\begin{affiliations}
Shachar Resisi, Yaron Bromberg\\
Racah Institute of Physics, The Hebrew University of Jerusalem, Jerusalem 91904, Israel\\
Email Address: yaron.bromberg@mail.huji.ac.il\\
\medskip
Sebastien M. Popoff\\
Institut Langevin, CNRS, ESPCI Paris, Université PSL, 75005 Paris, France

\end{affiliations}


\begin{abstract}
When multimode optical fibers are perturbed, the data that is transmitted through them is scrambled. This presents a major difficulty for many possible applications, such as multimode fiber based telecommunication and endoscopy. To overcome this challenge, a deep learning approach that generalizes over mechanical perturbations is presented.
Using this approach, successful reconstruction of the input images from intensity-only measurements of speckle patterns at the output of a 1.5 meter-long randomly perturbed multimode fiber is demonstrated. The model's success is explained by hidden correlations in the speckle of random fiber conformations. 
\end{abstract}


\section{Introduction}

Multimode optical fibers (MMFs) hold great promise for increasing the capacity of data transmission, especially for applications such as optical communication systems \cite{Richardson2013}, fiber lasers \cite{Wright2017} and endoscopic imaging \cite{Cizmar2012, Bianchi2012, Choi2012, CizmarInVivo2018, LiGuideStar2020}. 
An important challenge such applications face is the inherent sensitivity of fibers to various types of fluctuations, such as thermal, acoustic, or mechanical perturbations.
Unless special fibers are used \cite{Tsvirkun2019}, such perturbations dramatically change the transmission properties, since modal interference is extremely sensitive to changes in the phase accumulated by the fiber's guided modes, which are in turn affected by the perturbations. Overcoming the effects of these perturbations is an important step towards robust fiber-based technologies and applications. 

\hspace{0.2in}
For one static conformation of the fiber, the transmission properties are fully captured by the transmission matrix (TM) \cite{Sebastien2010PRL}. When weak perturbations are applied on the fiber, the TM of the deformed fiber can be predicted \cite{LiGuideStar2020, CizmarPRL2018}, mapped to pre-calibrated deformations \cite{FarahiMoserPsaltis2013}, or compensated for \cite{Caravaca-Aguirre2013, Ploschner2015, Loterie2017}. Unfortunately, these options do not hold for strong deformations. 
Invariant statistical properties can also be harnessed to recover the transmitted information.
For example, a rotational memory effect was observed in pixel space \cite{Amitonova:15}, and recently a similar effect in mode basis was used for image reconstruction through MMFs \cite{Li2020compressively}. 
Alas, these properties are also limited to small perturbations and require a prior estimation of the TM or a feedback signal. The existence of invariant properties that survive strong deformations would allow envisioning image reconstruction through unknown and strongly perturbed fibers. 

\hspace{0.2in}
The high availability and low cost of strong computing power in recent years gave a significant boost to deep learning (DL) approaches. Recently, neural networks have attracted increasing attention in the optical community, allowing for the reconstruction of input information after propagation through random complex media \cite{OpticaAnotherLi2018, LSARahmani2018, OpticaBorhani2018, OpticaLi2018, QueenMary2019, NaturalImages2019, DLReviewBarbastathis2019, Rahmani2020, Zhao2020, Chen2020, Zhu2020, Liu2020}. In fibers, convolutional neural networks (CNN) were shown to produce reconstructions with a similar fidelity to the TM approach \cite{LSARahmani2018, OpticaBorhani2018, QueenMary2019}. Most previous works were limited to a single, static, fiber conformation. It was recently shown that CNN models can reconstruct images from fibers that are weakly perturbed while the data sets were recorded. The weak perturbations were induced by natural drifts in the environmental conditions \cite{OpticaBorhani2018}, by weak bending of the fiber \cite{QueenMary2019, Liu2020} or by wavelength scanning \cite{Kakkava2020}.
Nonetheless, all previous works in fibers were not able to generalize to unknown and strongly perturbed fiber conformations that span a wide configuration space which describes numerous uncorrelated fiber configurations. 
DL approaches are known to efficiently learn invariant properties of signals, and can thus be harnessed for the challenge of learning the transmission through strongly perturbed systems. Indeed, in scattering media, Li \textit{et al.} trained a CNN on the speckle created by a group of thin diffusers, and produced excellent image reconstructions from speckle resulting from different diffusers of the same type \cite{OpticaLi2018, Li2020}. This generalization was possible due to the existence of correlations between speckle created by the different diffusers, an invariant property which the DL model learned to recognize.

\begin{figure*}[!ht]
    \centering\includegraphics[width=\linewidth]{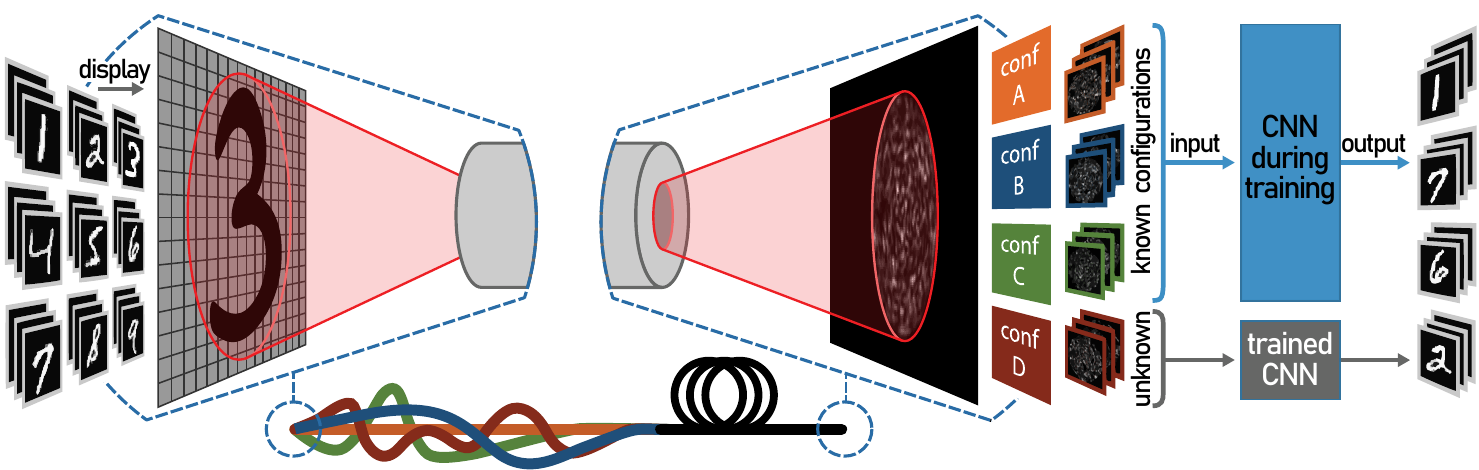}
    \caption{\textbf{Schematics of the reconstruction process.} Images from the MNIST hand-written digit dataset are displayed on a digital micromirror device (DMD). A laser beam is reflected from the DMD and injected on a multimode fiber, whose distal end is imaged on a CMOS camera. This procedure is repeated for each fiber conformation that we change by applying local bends at multiple positions along the fiber (depicted by the orange, blue, green and red curves). 
    For the same set of input images (hand-written digits), different configurations give different output speckle patterns. Speckle-digit pairs from some of the configurations (coined the known configurations) are used to train a convolutional neural network (CNN). Speckle from unknown configurations are used to reconstruct digit input images and estimate the generalization capabilities of the trained model.}
    \label{fig:gist}
\end{figure*}

\hspace{0.2in}
Motivated by these results, we use DL to learn invariant properties of strongly perturbed multimode fibers. We use a CNN and show that when we train the network over hundreds of random nearly uncorrelated fiber bends, it succeeds in reconstructing high-fidelity images even when the fiber is strongly perturbed many weeks after the training perturbations. We call this method of training on multiple low-correlated fiber conformations \textit{configuration training}.
A sketch of our workflow is presented as \textbf{Figure \ref{fig:gist}}. 
Configurations are created via strong mechanical perturbations, by simultaneously bending the fiber at multiple positions using an array of piezoelectric plate benders that are positioned above the fiber \cite{us}. 
We suggest that the generalization is possible due to some hidden statistical similarities in the speckle, and support this by showing these correlations along with a 2-dimensional (2D) embedding of the acquired speckle.

\section{Methods}

\subsection{Experimental setup}
\label{sec:setup}

The experimental setup, as depicted in \textbf{Figure \ref{fig:setup}}, consists of a HeNe CW laser (wavelength of $\lambda=632.8$ nm) which illuminates a digital micromirror device (DMD; Ajile AJD-4500-UT). The light from the DMD's \textit{on} pixels is imaged on the proximal end of a 1.5 meter-long step-index MMF (Thorlabs FG050LGA; V number $\approx55$ at $\lambda$) using a $4f$ system. The intensity of the output speckle pattern is imaged by a second $4f$ system on a CMOS camera. 
We place 37 piezoelectric actuators along the fiber, and use a computer to control their vertical displacement. Each actuator bends the fiber by a three-point contact, creating a bell-shaped local deformation of the fiber, and inducing mode mixing exhibited in the fiber’s transmission matrix \cite{Matthes2020}. The curvature of the bend depends on the vertical travel of the actuator, and is on the scale of millimeters (down to $\sim12$ mm; see Figure \ref{fig:setup}(b) and \ref{fig:setup}(c)) \cite{us}. By changing the actuator's position, and composing the bends created by all actuators, we can create a huge variety of possible configurations, with varying correlations between each other.
To quantify the correlation between different random fiber conformations, we compute the Pearson correlation coefficient (PCC) between speckle patterns obtained for a fixed input. When we randomize the bending configuration created by all of the actuators using their full stroke, the average PCC between different fiber conformations (calculated over the same input patterns) is $0.12\pm 0.01$, which we find to be equivalent to the PCC values obtained by simply bending the fiber on centimeter scales (see supplementary material for more details). These similar correlation values in multiple bending regimes emphasize the system's relevance for studying general bending deformations.

\subsection{Data acquisition and processing}
\label{sec:data}

To collect many different speckle patterns for a single fiber geometry, we display sequences of hand-written digit patterns from the MNIST dataset \cite{mnist} on the DMD and record the resulting $128\times128$ speckle pattern. Due to the binary nature of the DMD, each digit is first converted to a binary amplitude image by applying a threshold on the original 8-bit grayscale image. The number of digit patterns and fiber configurations we acquire varies in the different training approaches we study. In all cases, we acquire separate data for training and for testing purposes (according to the division of the original dataset \cite{mnist}). 

\begin{figure}[!ht]
    \centering\includegraphics[width=0.6\linewidth]{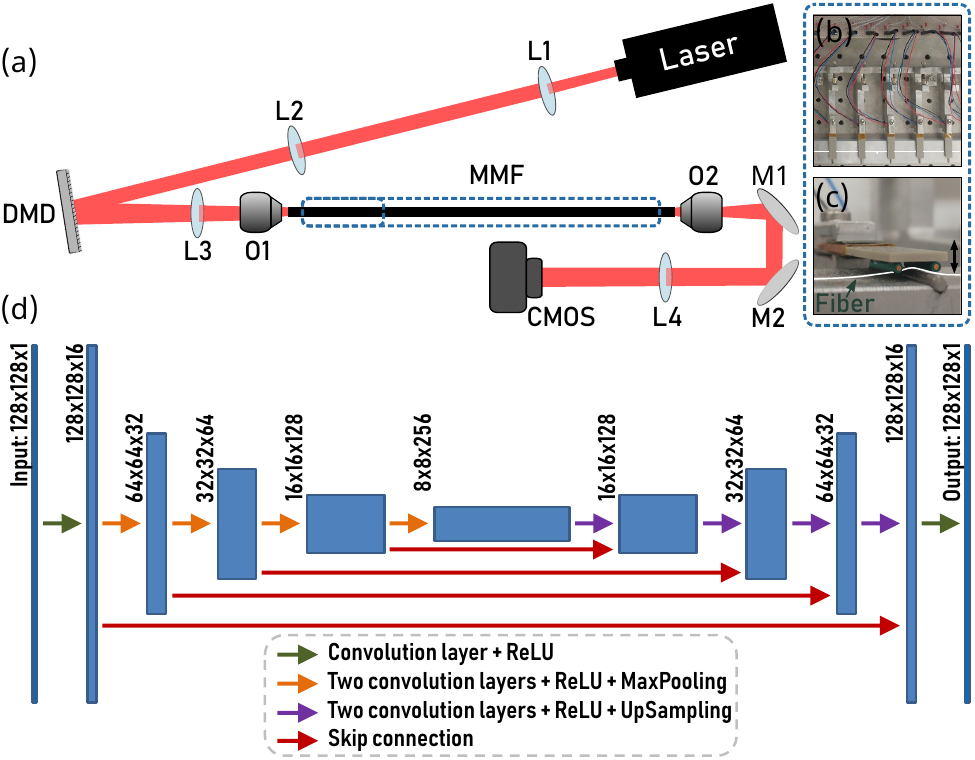}
    \caption{\textbf{Experimental setup and neural network architecture.} (a) A laser beam illuminates a DMD. The reflected light is imaged onto the proximal end of an MMF. Piezoelectric plate benders are positioned above the fiber (b), allowing for the application of various computer-controlled bends along the fiber. The light from the distal end of the fiber is imaged on a camera. (c) Side view of a bend, applied by a three-point contact between the actuator and an opposing metal rod. The curvature's radius varies on the scale of $mm$-s. (d) Schematic of the U-Net architecture used for image reconstruction from speckle obtained at multiple fiber geometries.
    L, lens; DMD, digital micromirror device; O, objective; MMF, multimode fiber; M, mirror; CMOS, camera.}
    \label{fig:setup}
    \label{fig:model}
\end{figure}

\hspace{0.2in}
The deep learning model we use is a convolutional neural network (CNN) of U-Net type \cite{unet}. We feed the network with speckle images, and reconstruct the digit patterns that were displayed on the DMD. Once the training is complete, predictions are made in real time (milliseconds). The exact architecture we use is depicted in Figure \ref{fig:model}(d) (see supplementary material for more details). 
The metrics we use to quantitatively appraise the performance of our model are the pixel-wise accuracy (defined as the percentage of the correctly predicted pixels) and the Jaccard index (JI; the intersection over union score of the binary reconstructions, which ranges between 0 to 1, and is only affected by the white pixels). Additionally, we train a very simple CNN \cite{myclassifier} to classify the reconstructions into digits, and compare each result with the digit number that was displayed on the DMD. We define the classification success as the true positive rate and calculate it over unknown patterns to assess the generalization capabilities of the trained model.

\section{Experimental Results}

\begin{figure}[!ht]
    \centering\includegraphics[width=0.6\linewidth]{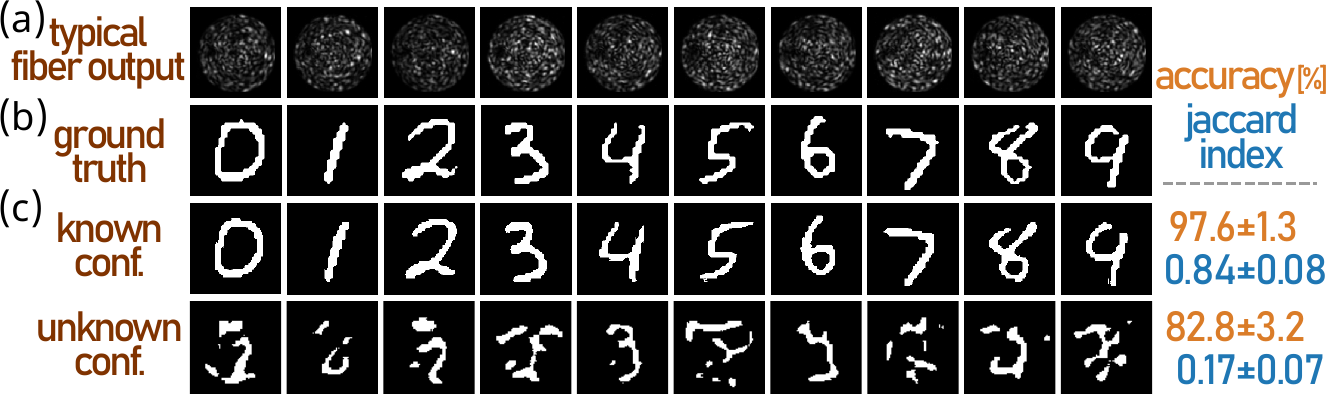}
    \caption{\textbf{Digit reconstruction using a CNN trained on a single fiber conformation.} 
    (a) Example of speckle patterns received at the distal end of the fiber for a random configuration. (b) The ground truth, i.e. the image that is sent to the DMD.
    (c) The top row shows representative examples of reconstructions from a known configuration (the one the model was trained on). The bottom row shows examples of an unknown configuration, from which it is evident that the model does not generalize to other fiber conformations. The corresponding inputs were not used as part of the training process. The pixel-wise accuracy, calculated over the patterns in each row appears in orange, and the Jaccard index (JI) in blue.
    }
    \label{fig:few}
\end{figure}

We start by demonstrating the reconstruction of a single configuration. For this first experiment we acquire a total of 70k different images (originating from 70k different hand-written digit patterns in the MNIST dataset), of which 60k are used for training and the rest 10k only for testing. 
We use the training set to train the model, and use the unknown test set to appraise its performance. 
As expected for an unperturbed fiber, and in accordance with previous works, the reconstruction is very accurate, see \textbf{Figure \ref{fig:few}}. Quantitatively, the average pixel-wise accuracy for the entire test set (averaged over 10k reconstructions) is $97.6\%\pm0.2$, the JI is $0.83\pm0.02$, and the classification success is over 97\%.
While yielding high fidelity reconstructions for the same fiber conformations, this model fails to generalize over unknown fiber perturbations, resulting in an average pixel accuracy of $82.8\%\pm3.2$ and JI of $0.17\pm0.07$ as demonstrated in the bottom row of Figure \ref{fig:few}(c).
 
\hspace{0.2in}
In principal, one could extend this approach and train a CNN on all of the data from multiple configurations. In the supplementary material, we demonstrate this for simultaneously training on 8 random fiber conformations. However, since each actuator induces significant mixing between the fiber modes \cite{Matthes2020}, the space representing the possible configurations induced by 37 actuators is very large, even for a short fiber. To statistically explore this space, a large number of configurations needs to be represented in the training set. The current approach, to which we refer as \textit{standard training}, is not scalable for a large number of configurations, as it requires a large data set for each configuration and thus the training set becomes too large to be handled efficiently in terms of memory (needed to store the images) and time (to train the network).

\hspace{0.2in}
As one cannot expect to learn all of the possible conformations of the fiber, predicting the output from an unknown configuration can be possible if there are invariant properties that are robust to conformation variations and are learned by the CNN. To harness these potential invariant properties, 
we train the network over 943 fiber conformations, obtained by randomizing the positions of all of the actuators. The degree of correlation between fiber conformations, quantified by the PCC of speckle patterns obtained for the same fiber input, is $0.12\pm0.01$. To account for the large number of configurations while limiting the size of the training set, we use only 800 training images per configuration and record the intensity of the resulting speckle. In total, data was acquired over the span of 14 weeks, during which a few different macro bends were applied in addition to the actuator-induced bends to improve the model’s robustness to mechanical perturbations of varying scale. The same average PCC was obtained between configurations from the same and different days, regardless of the applied macro bend. We acquire additional test data from 800 other random fiber conformations, to appraise the performance of the model on unknown configurations.
We coin this type of training, which consists of less data from multiple fiber conformations as \textit{configuration training}, because we prompt the model to learn general statistically invariant properties.

\begin{figure}[!ht]
    \centering\includegraphics[width=0.6\linewidth]{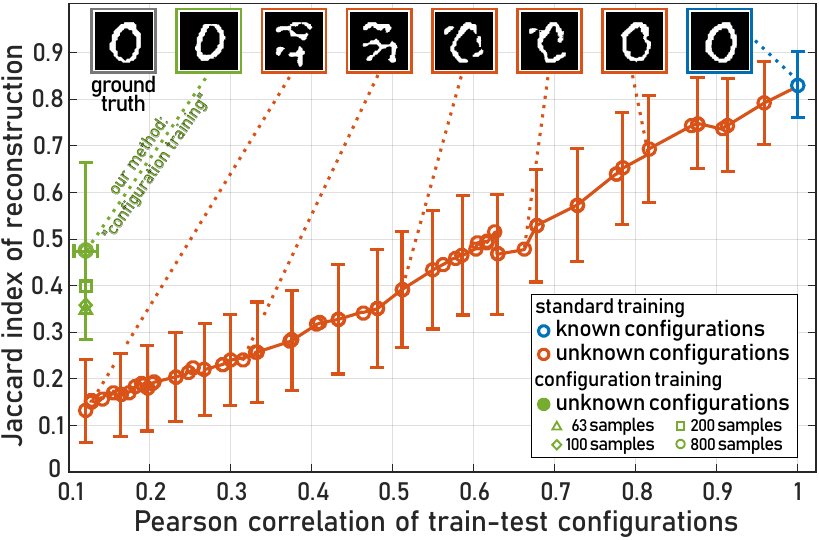}
    \caption{\textbf{Reconstruction fidelity vs. the correlation between the test configurations and the training set configurations.}
    The Jaccard index (JI), which quantifies the reconstruction fidelity, is plotted against the Pearson correlation coefficient (PCC) between the train and test configurations. For configurations with an average PCC of 0.12, our method of \textit{configuration training} (green circle) produces higher-quality reconstructions than standard (single configuration) training, by a factor of $\approx3.5$.
    The blue data point represents the JI for using the same fiber configuration for training and testing. 
    The red curve shows the degradation of reconstruction fidelity with the decreasing correlation between the train and test set (see supplementary material). The green data points correspond to unknown fiber conformations using \textit{configuration training} on 943 random fiber conformations. The effect of the training set size is observed by training separate models on 63/100/200/800 samples per configuration, with the average obtained JI depicted as the green triangle/diamond/square/circle. The standard deviation of the first three is emitted for clarity, and has the same size as the one that is shown. All points are calculated over the same test input patterns. 
    The PCC between two different configurations corresponds to the average PCC between respective output intensity patterns for the same input excitations, and the standard deviation between the train and test for \textit{configuration training} is shown. A representative example for the reconstruction of an unknown input digit is displayed at the top along with its ground truth.}
    \label{fig:jipcc}
\end{figure}

\hspace{0.2in}
The configuration training immensely improves the reconstruction of test images from unknown fiber conformations. We observe that when the average correlation between configurations from the train set and the test set (calculated for the same input patterns) is 0.12, the average Jaccard index increases from $JI=0.13$ for standard training to $JI=0.47$ for configuration training.
To emphasize the performance difference between our configuration training and the standard (single configuration) training, we plot the JI for standard training as a function of the output intensity correlation between the test configurations and the train set (\textbf{Figure \ref{fig:jipcc}}, red curve). The test configurations range from strong perturbations (train-test correlation $\sim0.12$) to weak perturbations (train-test correlation $\leq 1$). Thus, with our configuration training, reconstructions from strong perturbations have similar fidelity to ones that are achieved from weak perturbations with standard training. 
Noticeably, when the train-test correlation is $\sim0.6$ (weak perturbations regime), the JI values of standard training are similar to those obtained with our configuration training at a correlation of 0.12 (strong perturbations regime).
As depicted by the example reconstruction of Figure \ref{fig:jipcc}, this improvement translates an unintelligible image (leftmost red frame) to a sharp image (green) which greatly resembles the ground truth. Additional reconstruction examples are provided in Visualization 1 as part of the supplementary material. To study the impact of the size of the training set in this approach, we trained three additional models, where instead of 800 samples per configurations we used only 63/100/200 samples from each of the 943 configurations. We then tested the reconstruction fidelity using these models over the same test patterns of unknown configurations. The obtained average JI is depicted in Figure \ref{fig:jipcc} as the green triangle/diamond/square (correspondingly). Noticeably, the generated reconstructions have a lower JI than when the model was trained on 800 samples per configurations, however there is still an increase of a factor of approximately 3 to the average JI compared with the “standard training” approach, with a training set of comparable size and the same average configuration PCC.

\begin{figure}[!ht]
    \centering\includegraphics[width=0.5\linewidth]{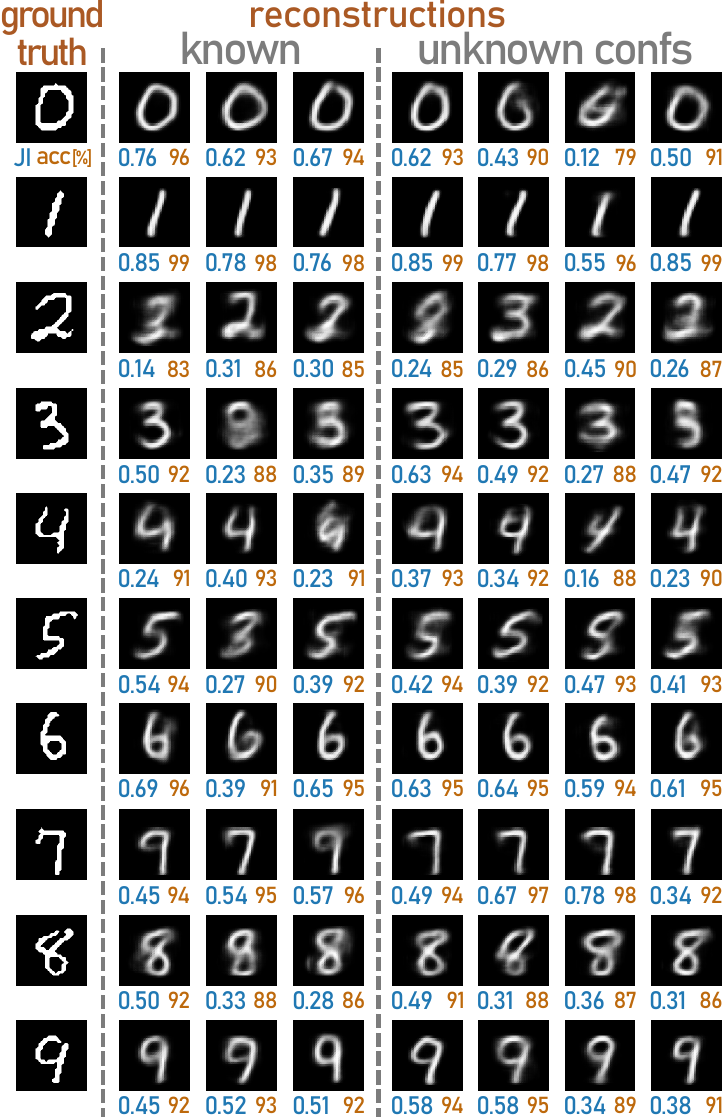}
    \caption{\textbf{Reconstructions produced by \textit{configuration training} of speckle from multiple low-correlated fiber conformations.}
    The leftmost column shows the ground truth. Each of the other columns show reconstructions from a randomly chosen configuration, acquired at different days. The three (four) columns under “known” ("unknown confs") depict reconstructions from configurations the model did (did not) train on. Noticeably, both known and unknown configurations results in reconstructions of similar fidelity. The Jaccard index of each reconstruction appears below it in blue, next to the pixel-wise accuracy which appears in orange.}
    \label{fig:many}
\end{figure}

\hspace{0.2in}
To further examine the configuration training results, we show representative examples of test patterns in \textbf{Figure \ref{fig:many}}, and in Visualization 2. In Figure \ref{fig:many}, each column describes reconstructions from an arbitrary configuration, one from each day the data was acquired.
Interestingly, the reconstructions for both known (part of the training) and unknown (only used for testing) configurations give results of similar quality. This is reflected by similar values which are obtained for known and unknown configurations over the entire test set using all of our evaluation metrics, as detailed in Table 1 of the supplementary material.
We attribute the resemblance of the results for known and unknown configurations to the small number of examples in the train set from each configuration. 
More training data from each of the configurations could produce even better reconstructions.
Moreover, the large standard deviation of the JI depicted in Figure \ref{fig:jipcc} by the error bars is manifested in Figure \ref{fig:many}, where evidently some digits are easier to reconstruct than others (e.g. '1' and '6' compared with than '2' and '3').

\section{Discussion}

\begin{figure}[!ht]
    \centering\includegraphics[width=0.6\linewidth]{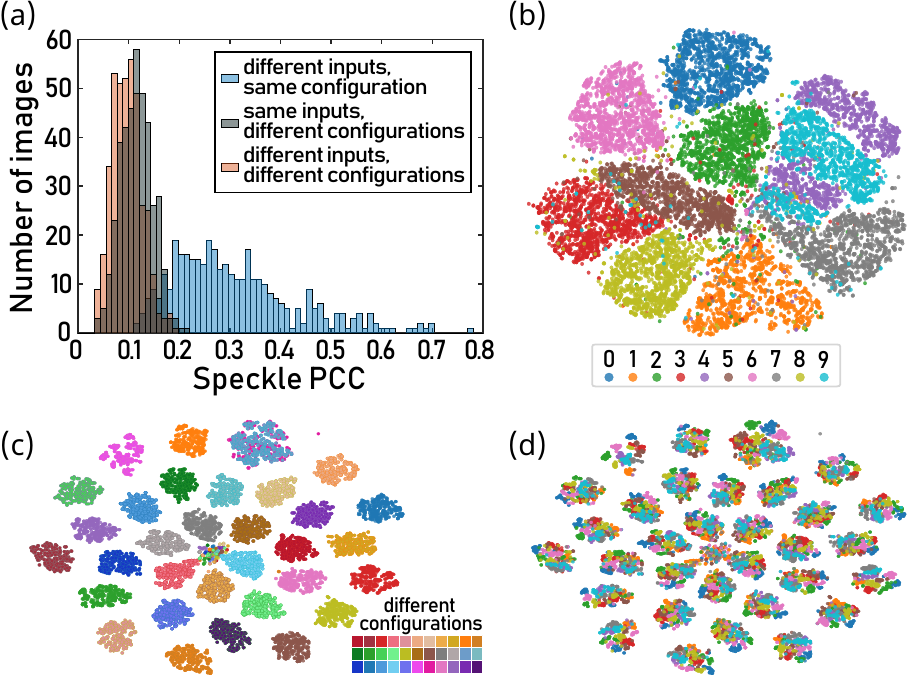}
    \caption{\textbf{Speckle correlations and 2D embedding of inputs and configurations.} 
    (a) Histogram of the Pearson correlation coefficients of speckle for different DMD inputs within the same configuration (blue), speckle generated for the same DMD inputs in two different configurations (gray) and different inputs in two different configurations (orange). The blue histogram is wide and accounts for the ease of reconstructing images from a single fiber conformation. The shift of the mean value between the gray and orange histograms could account to the reason a CNN is able to generalize over fiber bends.
    (b) 2D embedding of 15k speckle patterns from a single fiber conformation mostly divides them according to the class of the digit that was displayed on the DMD (0-9).
    (c-d) 2D embedding of speckle patterns from 33 random configurations. The embedding contains 32 blobs, and an additional centered blob that corresponds to data that was not easily separable (which appears in the center due to the algorithm's working mechanism). 
    (c) The color code corresponds to different configurations (information that was not available to the algorithm), showing configuration clusters (with two configurations that completely overlap). (d) Same data as (c), the colors correspond to digits (same code as (b)), and shows that t-SNE is not successful in separating this data according to digits.}
    \label{fig:stat}
\end{figure}

For a static fiber conformation accurate results can be obtained using a CNN with standard training, since similar digit images (e.g. two patterns of the digit '9') excite similar fiber modes. Thus, the resulting speckle for similar inputs within the same fiber conformation are spatially correlated, as shown in the blue histogram of \textbf{Figure \ref{fig:stat}}(a) for an input image of a '9'. These correlations aid the CNN to produce accurate reconstructions, as evident in the top row of Figure \ref{fig:few}(c) and the high JI described by the blue point of Figure \ref{fig:jipcc}. Furthermore, the relatively high correlations in the pixel basis hint that the simpler task of classifying the input digit images (according to digit) can potentially be achieved with a "classical" machine learning approach, i.e. without DL. 
In \cite{Li2020}, Li \textit{et al.} used an unsupervised dimension reduction technique and demonstrated that speckle that emerge from thin diffusers can be clustered according to their original class or acquisition configuration. Here we take a similar approach and show that using the t-distributed Stochastic Neighbor Embedding (t-SNE) dimensionality reduction technique \cite{tSNEVanDerMaaten2008}, images from the same fiber conformation are mostly clustered according to their underlying digit class (Figure \ref{fig:stat}(b)).

\hspace{0.2in}
For a dynamic fiber that undergoes strong perturbations, one would not expect the CNN to work since even for the same images at the fiber input, speckle patterns for different fiber deformations show low correlations. 
However, the transmission properties cannot be totally uncorrelated, as it would have prevented the generalization over unknown conformations that we experimentally demonstrated in this study. To explore how we are able to produce high fidelity reconstructions, we search for invariant properties in the speckle produced for different fiber conformations. Following \cite{OpticaLi2018}, we computed the correlations between output speckles for different configurations.
For two random configurations, the speckle patterns that are obtained for different inputs exhibit a much narrower correlation distribution, which is centered around a much smaller value than within the same fiber conformation, as we depict in the orange histogram of Figure \ref{fig:stat}(a). This distribution is centered at $0.11\pm0.01$, and presumably deems the mission of reconstruction from different configuration impossible.
However, when we compare between the same inputs in random configurations, this distribution is slightly shifted to higher values, and centered at $0.12\pm0.01$ (gray histogram of Figure \ref{fig:stat}(a)). 

\hspace{0.2in}
The observed decorrelation of the output intensity pattern in the pixel basis does not allow us to directly assess the level of disorder, since even if the transmission speckle decorrelates quickly when deformation is applied, some hidden information can still be present \cite{Ploschner2015, CizmarPRL2018}.
We believe that the mere existence of these low, but non-zero correlations stand at the heart of the CNN's success in the image reconstruction through unknown fiber conformations.
Indeed, the 2D embedding of speckle patterns from different inputs and configurations using the t-SNE algorithm clusters the data points according to their original configuration. In Figure \ref{fig:stat}(c) we show a representative example for the embedding of 33 random configurations (with two configurations that completely overlap).
We note that for less data, t-SNE is unable to differentiate between digits (Figure \ref{fig:stat}(d)) - a task the CNN succeeds at. 
The fact that configurations of the disorder can be discriminated using statistical analysis tools (Figure \ref{fig:stat}(c) and \ref{fig:stat}(d)) shows that the transmission properties of light are not totally randomized by the deformations, as outputs would otherwise be indistinguishable. This supports the interpretation that the deep learning model learns the invariant properties, common to all deformations, which are then used to infer and reconstruct the input excitation.
We therefore conclude that there is an advantage to using DL when dealing with multiple conformations.

\hspace{0.2in}
The fiber conformations we study in this work are driven by actuators that induce multiple bends on the millimeter scale. In real life applications, we expect fewer bends though on a longer length scale. We therefore compare the correlation between different fiber conformations, obtained for a few macro-bends and for the actuator-driven bends, and find that the PCC of the speckle output is $\approx 0.12$ in both cases (see supplementary material for more details).
We therefore believe that the proposed configuration training can be implemented in real-life applications, and in particular for biomedical imaging, since the parameters of the fiber we use adhere to those used in prototype microendoscopes \cite{CizmarInVivo2018}.
A note-worthy limitation of our work emerges from the homogeneity of the train set, which consists of a single class of images: digits. The CNN we use channels the input through an encoder, which represents the data in a smaller dimension called the latent space. Due to the homogeneity and the nature of encoders, code in the latent space would be decoded to a digit, or to a composition of digits, in the pixel space (the reconstruction). Thus, images from the same domain that is to be reconstructed should be used for the training of our model to achieve the best performance. However, this shortcoming can be easily overcome by utilizing a more complex DL model, which was shown to be able to generalize to images from other domains \cite{LSARahmani2018, OpticaLi2018}.

\section{Conclusion}
In this work we experimentally demonstrated robustness to fiber deformations using a deep learning model. Our model is able to reconstruct images which are transmitted through a bent multimode fiber, with no knowledge of the specific fiber conformation, and regardless of the inflicted bend configuration. 
We presented an immense improvement in reconstruction fidelity compared with standard DL approaches, transforming an incomprehensible image to an intelligible one.
We showed that to achieve generalization it is essential to train a CNN on the speckle patterns that originate from numerous fiber bends and are acquired for many different image inputs. We implemented such an approach by introducing \textit{configuration training}. We then tested the network's performance on patterns from strongly perturbed fibers, and showed good reconstruction results even for strong perturbations applied many weeks after the training process.
Our demonstration has possible real-life applications, in particular for fiber endoscopy where changes in the geometrical configuration of the fiber are unavoidable. The CNN model we used is fairly simple and compact, with low memory requirements (compared to previous works in the field), allowing for video rate implementation with standard computers.

\medskip
\textbf{Supporting Information} \par 
See Supplement 1 for supporting content. See Visualization 1 for a few examples that provide a more extensive view of Figure \ref{fig:jipcc}, and Visualization 2 for more reconstruction examples using our configuration training.
Supporting Information is available from the Wiley Online Library or from the author.

\medskip
\textbf{Acknowledgements} \par 
The authors kindly thank Snir Gazit for providing access to the computational resources which were used to train the neural networks, along with Ori Katz and Roy Friedman for many fruitful discussions and suggestions. 
SR and YB acknowledge the support of the Israeli Ministry of Science and Technology and the Zuckerman STEM Leadership Program. SMP is supported by the French \textit{Agence Nationale pour la Recherche} (grant No. ANR-16-CE25-0008-01 MOLOTOF), the Labex WIFI (ANR-10-LABX-24, ANR-10-IDEX-0001-02 PSL*) and the France's \textit{Centre National de la Recherche Scientifique} (CNRS; France-Israel grant PRC1672). All authors acknowledge the support of \textit{Laboratoire international associé Imaginano}.

\medskip

%
\bibliographystyle{MSP}
\bibliography{refs}

\end{document}


\maketitle

\section{System stability}
For a constant input on the DMD, the output speckle intensity remains very highly correlated (Pearson correlation coefficient $>0.99$ between the output speckle at time $t$ and the speckle pattern acquired at $t=0$) over the course of 17 hours. This stability period is longer than the time it takes to acquire data from a fiber conformation, which is approximately 1 hour, by more than an order of magnitude. When the same fiber configuration is repeated after 7 weeks (during which other configurations were applied), the correlation decreases to $0.61\pm0.04$ due to a combination of wavelength, temperature and mechanical drifts.


\section{Comparison of actuator-induced bends and macro-bends}
\begin{figure}[!ht]
    \centering\includegraphics[width=0.7\linewidth]{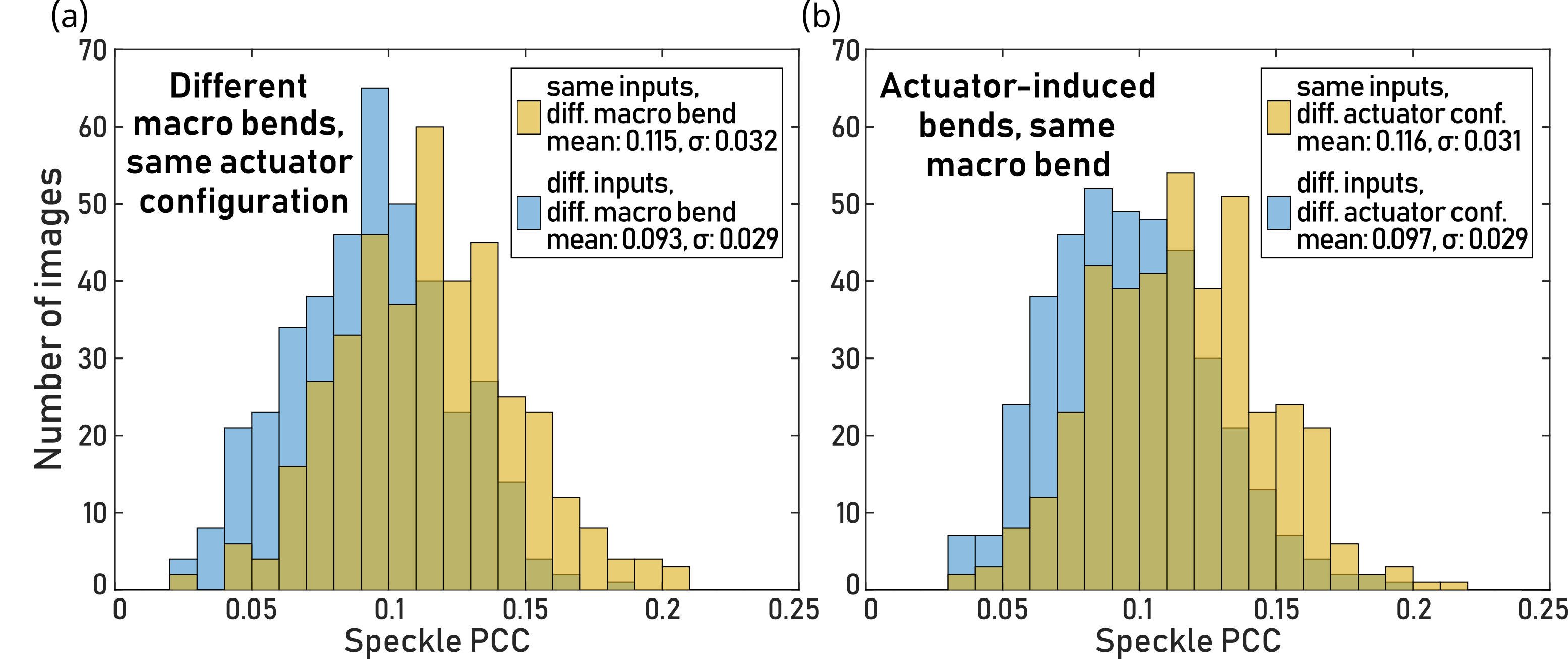}
    \caption{\textbf{Speckle statistics in different macro-bends and comparison to actuator-induced bends.}
    (a) Histogram of the Pearson correlations of speckle patterns for the same DMD inputs in different macro-bends conformations (orange), and speckle generated for different DMD inputs in two macro-bends (orange). The average value and standard deviations of the two distributions are very close to those obtained for actuator-induced bends, as depicted in Fig. 6(a) of the main text and reproduced in (b).}
    \label{fig:supstat}
\end{figure}

As we detail in the experimental setup section of the main text, our system is comprised of an array of piezoelectric plate bender actuators, which are positioned above the fiber. These actuators bend multiple segments of the fiber (with a curvature radius down to approximately 12 mm), to create a specific fiber conformation. Two such configurations have an average Pearson correlation coefficient (PCC) of $0.12\pm 0.01$, which we calculate based on the intensity of the speckle patterns for the same input. The PCC is calculated only over the fiber core, to eliminate the effect of the pixels which surround it in the acquired squared image.
To compare between the change of correlation created by our actuator-induced bends and macro-bends, which are also relevant for real-life applications, we calculate the PCC between different macro-bends in the same manner. We perform six different macro-bends along the fiber (i.e. bend the fiber on the centimeter scale at different angles and positions), display various inputs on the DMD and acquire the resulting speckle patterns. We observe that the correlation between the speckle for these macro-bends behaves similarly to the actuator-induced bends. Specifically, we see that on average the PCC between speckle patterns for the same inputs in different macro-bends is also 0.12, as we show in the orange histogram of Fig. \ref{fig:supstat}(a). Moreover, the correlation distribution for different inputs in different macro-bends is again slightly shifted relative to the same inputs, similarly to the shift we observe for actuator-induced conformations (see the discussion and Fig. 6 of the main text). From this similarity, as we depict in Fig. \ref{fig:supstat}, we conclude that actuator-induced bends simulate the effects of macro-bends on the intensity of the speckle pattern at the distal end of the fiber accurately, which allows us to use actuator-induced bends to study bend robustness in MMF.

\section{Neural network architecture}
To reconstruct the DMD inputs from the speckle intensity patterns, we use a neural network. Out of the wide variety of architectures that were explored in previous works for related purposes, including fully-connected and complex models \cite{trans, actor, simple}, we choose a standard "U-Net" type convolutional neural network (CNN) \cite{unet}, which we implement using the Keras and TensorFlow python libraries. This model embraces the encoder-decoder structure, along with "skip" connections between corresponding layers in the encoder and the decoder, to capture both global and local features in the images. As depicted in Fig. 2(d) of the main text, our network consists of all-convolution blocks. A block is built from two convolution layers with the same convolution kernels sizes, each followed by the rectified linear unit (ReLU) activation function. Each block is followed by a max pooling layer in the encoder or an up sampling layer in the decoder. The output is then received by applying another single convolution layer with a $1\times1$ kernel, again followed with the ReLU activation. We found that this model produces high fidelity reconstructions, while having a relatively small number of learnable parameters (less than 2 million) due to its efficient all-convolution nature.
To train the model, we use the Adam optimizer with an initial learning rate of $10^{-4}$, and a decay factor of 0.5 every 10 epochs. 
We feed the network a $128\times128\times1$ speckle pattern (8-bit grayscale image) and receive as output a $128\times128\times1$ grayscale digit reconstruction, which we convert to binary using a constant threshold. Using the binary cross entropy loss function we compare the generated reconstruction with the binary image that was displayed on the DMD. After the training is complete, predictions are made in real time.

\section{Learning a few random configurations}
\begin{figure}[!ht]
    \centering\includegraphics[width=0.7\linewidth]{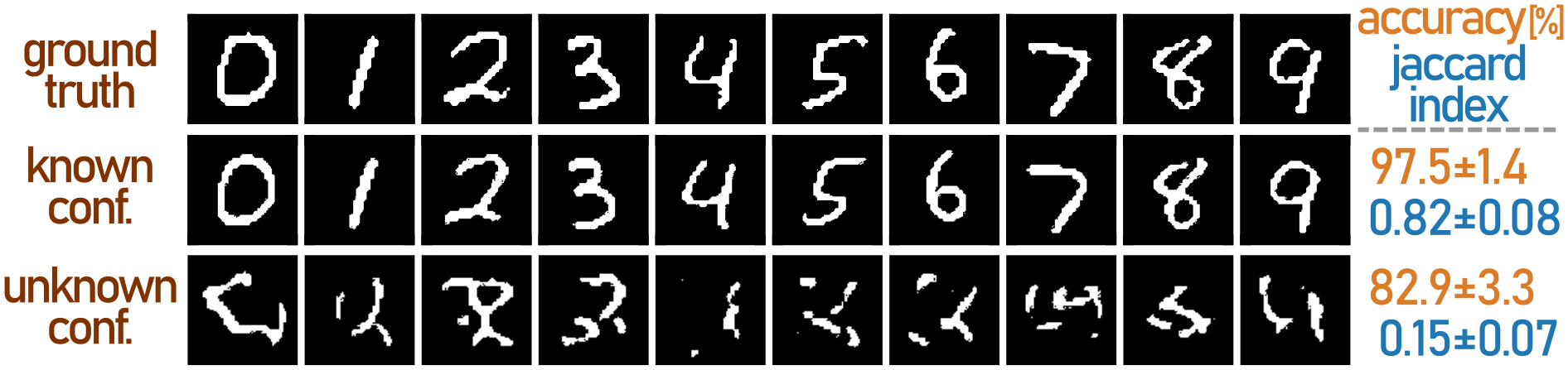}
    \caption{\textbf{Digit reconstruction using a CNN trained on eight configurations.}
    Representative reconstructions produced by a model that was trained on the full train set of eight random fiber conformations (standard training). The reconstructions in the second row originate from a known configuration (part of the train set), while the bottom one is from an unknown, and low-correlated fiber conformation. Both should be compared with the ground truth (top row). The corresponding inputs were not used as part of the training. The pixel-wise accuracy, calculated over the patterns in each row appears in orange, and the Jaccard index in blue.}
    \label{fig:supfew}
\end{figure}

Once we establish that our model reproduces known results by reconstructing the input at the proximal end of a MMF using the output speckle from a single fiber conformation (Fig. 3 of the main text), we wish to investigate if the CNN has the possibility of overcoming perturbations. For this purpose, we first have to ensure it can reconstruct data from several, low-correlated fiber bends. We therefore train a CNN on eight different, randomly created, fiber conformations (corresponding to eight different actuator-induced bends of the fiber). The average Pearson correlation between speckle patterns from these configurations (resulting from the same input) is $0.14\pm0.05$. 
We record the speckle patterns resulting from projecting the MNIST dataset on the DMD for the eight configurations. Thus, the training set consists of 480k speckle-digit pairs. 
In Fig. \ref{fig:supfew} we show representative examples for reconstructions of 0-9 digits for one of these eight configurations. Compared with the ground truth, the predictions have high fidelity, though subtle differences are visible. These reflect as a slightly lower accuracy and JI than those found with a model that was trained on a single fiber conformation, as discussed in the main text. Over the entire test set (averaged over all configurations) the accuracy is $97.4\%\pm0.2$, the JI is $0.81\pm0.03$ and the classification success remains over 97\%. Reconstructions of speckle patterns from other configurations using this trained model yield poor results, as evident by the bottom row of Fig. \ref{fig:supfew}. The network’s failure to generalize to an unknown fiber configuration illustrates the high level of complexity of the applied perturbations, which results from mode mixing and modal interference induced by a random configuration of 37 actuators.

We therefore see that using a rather simple CNN model, to which we provide a large training set for each of the configurations (i.e. using the standard training approach), we can achieve good reconstructions of known configurations.
However, as we mention in the main text, doing so is not scalable for a large number of configurations, as the training set becomes too large to be handled efficiently. We note that already for the eight configurations the training set consists of nearly half a million image pairs. 

\section{Reconstruction degradation with decreased PCC to the train set}
As we have shown in the main text, when a model is trained on a large training set from a specific fiber conformation, predicting the input of an output speckle from a configuration with a high PCC to the train set yields reconstructions with a very high fidelity and Jaccard index (JI). When the fiber is then perturbed, the correlation of the emerging configuration with the training configuration becomes smaller. This change of correlation is larger (i.e. smaller correlation) as the perturbation gets stronger. The decrease of correlation leads to a degradation in the reconstruction fidelity. The stronger the perturbation, the less the generated output resembles the ground truth, and therefore the reconstruction's JI is also smaller. In the red curve of Fig. 4 in the main text we show this gradual change. To create configurations whose difference from a specific fiber conformation varies on a large scale (x axis) we composed multiple weak perturbations on the original configuration, and acquired image pairs in each of these new configurations. We then used a model that was trained on the original conformation to create image reconstructions from the speckle patterns from all of the perturbed configurations, and noticed the almost linear trend depicted by the red curve.

\section{Learning multiple weak perturbations}
To limit the size of the training set while working with an increasing number of fiber bends, we start acquiring much less image pairs for each configuration. To see if the model successfully learns invariant statistical properties of the physical system, we begin with a random fiber conformation, and acquire data. We then perform a 'step', by inflicting a small change to the curvature of the bend a random actuator applies, acquire more data, and repeat this process for 300 steps. 
For each of the steps we display different train digits and the same test digits on the DMD and record the intensity of the resulting speckle. The PCC between speckle generated for the same inputs in adjacent steps (i.e. before and after a weak perturbation) is always larger than 0.7.

\begin{figure}[!ht]
    \centering\includegraphics[width=0.7\linewidth]{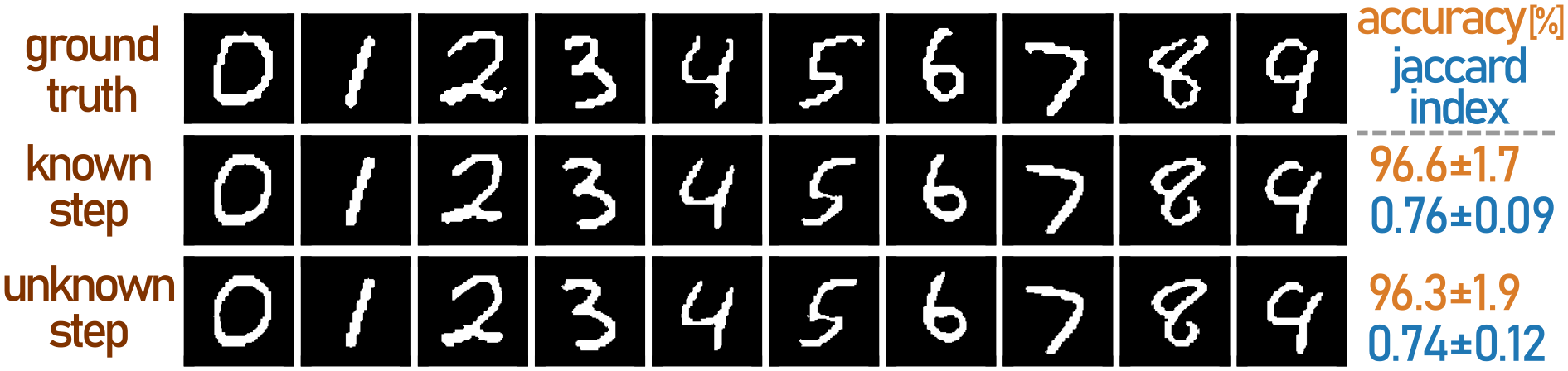}
    \caption{\textbf{Digit reconstruction using a CNN trained with \textit{configuration training} on multiple weak perturbations.}
    Representative reconstructions produced by a model that was trained on little data from multiple fiber conformations, differing by weak mechanical perturbations. The reconstructions in the second row originate from a known configuration (part of the train set), while the bottom one is from an unknown, but correlated, configuration. Both should be compared with the ground truth (top row). The corresponding inputs were not used as part of the training. The pixel-wise accuracy, calculated over the patterns in each row appears in orange, and the Jaccard index in blue.}
    \label{fig:supweak}
\end{figure}

First, we train a CNN on half of the data, by choosing the first five steps out of every ten, and test the reconstruction. We find that the reconstruction has high fidelity, both for the known and similar though unknown configurations. Representative examples of the reconstructions of both cases are shown in Fig. \ref{fig:supweak}. Interestingly, the reconstruction has a similar fidelity for both known and unknown configurations, as reflected by the accuracy and the JI. We attribute this to the small amount of data from each step that we used for training, which results in the lack of over-fitting by the model, and to the high correlations between adjacent steps.
To witness the effect of larger perturbations, We induce larger curvature differences between consecutive steps. The PCC between adjacent steps here is larger than 0.55. When we train the CNN on this data, we see that it produces reconstructions which are qualitatively and quantitatively similar to the case of weaker perturbations. We therefore deduce that even with less data, the CNN is able to generalize over weak perturbations, whose correlations to known configurations is larger than 0.55. 

\begin{figure}[!ht]
    \centering\includegraphics[width=0.5\linewidth]{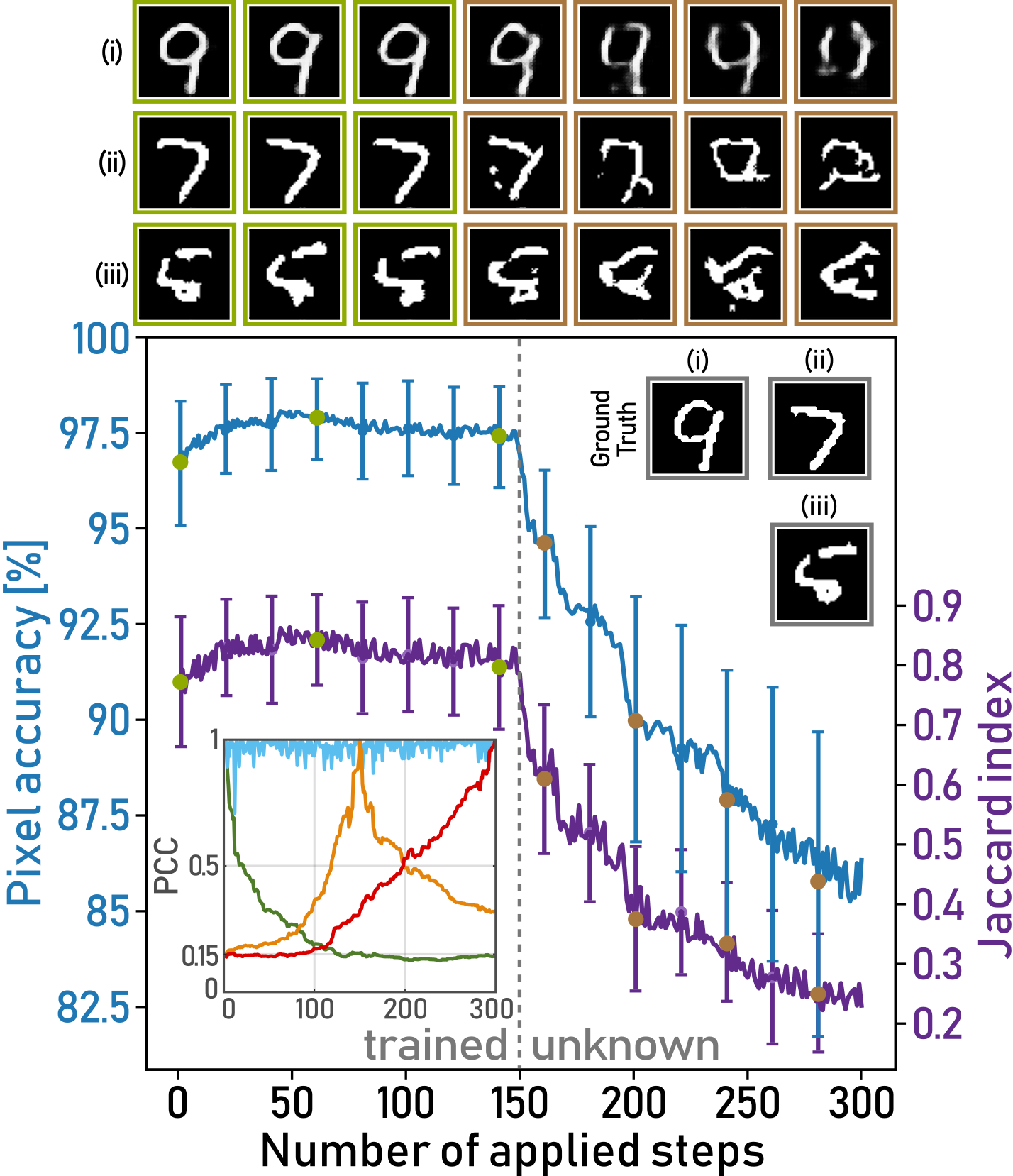}
    \caption{\textbf{Learning multiple consecutive fiber configurations with weak perturbations.}
    Average pixel-wise accuracy (blue curve) and JI (purple curve) for digit reconstruction using a CNN trained on data from (the first) 150 fiber configurations. Configurations vary by a weak perturbation, with a correlation $>0.7$ between consecutive steps.
    Pearson correlation coefficients (PCC) of the acquired speckle intensities are displayed in the bottom inset. The sky-blue curve depicts the correlation between consecutive configurations, while the green (/red/orange) curve shows the correlation between each configuration and the first (/last/middle) one.
    (i-iii) reconstruction examples of test patterns, which should be compared with the corresponding ground truth. The first three reconstructions in each row (framed in green) are from trained configurations (identifiable by green circles on the accuracy/JI curves), while the other four (framed in brown) are from configurations that were not trained on (identifiable by brown circles on the main curves), and show the prediction degradation as the source configuration becomes less correlated with the training configurations. The predictions in (i) ((iii)) correspond to the upper (lower) part of the standard deviations, and those in (ii) are close to the average values.}
    \label{fig:supsimilar}
\end{figure}

Next, to try and improve the robustness of the model to weak perturbations, we train a CNN on the first 150 out of the 300 acquired steps and test the accuracy and JI of the reconstructions from all steps. We witness an almost constant performance for all known configurations and a degradation in the reconstruction fidelity as the configuration's similarity to the training set decreases (larger step numbers). 
These findings are shown in Fig. \ref{fig:supsimilar}, along with the configurations correlations which are calculated over the same DMD inputs, and three reconstruction examples. 

\section{Reconstruction evaluation}

\begin{figure}[!ht]
    \centering\includegraphics[width=0.6\linewidth]{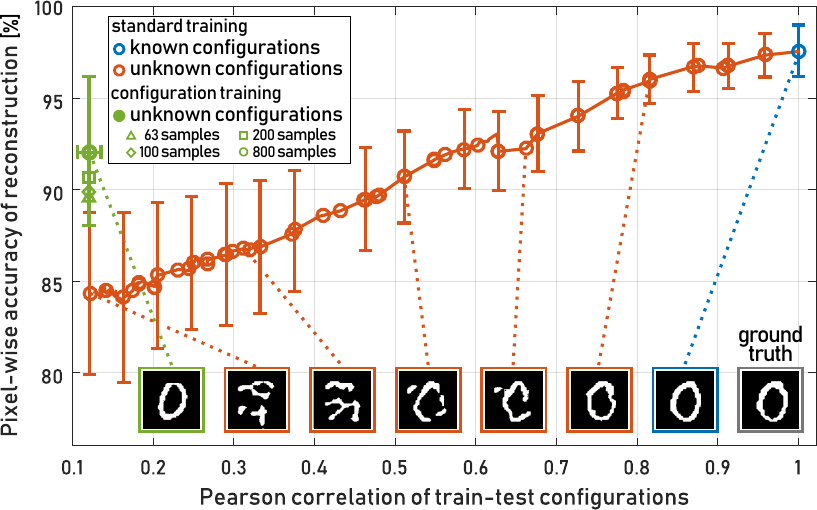}
    \caption{\textbf{Reconstruction accuracy vs. the correlation between the test configurations and the training set configurations.}
    The pixel-wise accuracy is plotted against the Pearson correlation coefficient (PCC) between the train and test configurations, similarly to Fig. 4 of the main text where the Jaccard index is depicted.}
    \label{fig:supaccpcc}
\end{figure}

In Fig. \ref{fig:supaccpcc}, using the same data as in Fig. 4 of the main text, we depict the pixel-wise accuracy of the reconstructions that are generated for different perturbations of varying potency. We witness the same trend that is shown in Fig. 4, where the reconstruction fidelity decreases with lower correlations and increases using our configuration training approach. 
Table \ref{tab:supacc} compares between the results we achieved for the different experiments in the three metrics we used. In all experiments, all metrics are calculated over test patterns, which are unknown to the trained model (were not used as part of the training). The second line, whose name reads "Standard training, one configuration" refers to the original experiment, where a CNN is trained on a single, static, fiber conformation. "Standard training, few configurations" refers to the experiment where we trained a CNN on eight random fiber configurations, whose correlations are approximately 0.14, as presented above. Next, the line that is denoted as "Configuration training, weak perturbations" includes the metrics values which are calculated over the consecutive 150 trained steps, as discussed above and shown in Fig. \ref{fig:supsimilar}. Lastly, the first row of "Configuration training, strong perturbations" is calculated over the 943 configurations the model trained on, while its bottom row is calculated over the 800 unknown (randomly created, low-correlated) fiber conformations that were used for testing only.

\begin{table}[ht]
    \centering
    \resizebox{0.7\textwidth}{!}{
    \begin{tabular}{|c|c|c|c|c|}
    \hline \textbf{Experiment} & \textbf{\begin{tabular}[c]{@{}c@{}}Relevant\\ figure\end{tabular}} & \textbf{\begin{tabular}[c]{@{}c@{}}Pixel\\ accuracy [\%]\end{tabular}} & \textbf{\begin{tabular}[c]{@{}c@{}}Jaccard\\ index\end{tabular}} & \textbf{\begin{tabular}[c]{@{}c@{}}Classification\\ error [\%]\end{tabular}} \\ \hline
    \begin{tabular}[c]{@{}c@{}}\textit{Standard training},\\ one configuration\end{tabular} & 3 & $97.6\pm0.2$ & $0.83\pm0.02$ & $2.56\pm1.47$ \\ \hline
    \begin{tabular}[c]{@{}c@{}}\textit{Standard training},\\ few configurations\end{tabular} & S2 & $97.4\pm0.2$ & $0.81\pm0.03$ & $2.63\pm1.58$ \\ \hline
    \begin{tabular}[c]{@{}c@{}}\textit{Configuration training},\\ weak perturbations\end{tabular} & S4 & $97.6\pm1.0$ & $0.82\pm0.09$ & $2.76\pm1.28$ \\ \hline
    \multirow{1}{*}{\begin{tabular}[c]{@{}c@{}}\textit{Configuration training},\end{tabular}} & \multirow{2}{*}{4, 5} & $92.0\pm3.8$ & $0.50\pm0.20$ & $23.02\pm0.04$ \\ \cline{3-5}  strong perturbations &  & $91.6\pm4.0$ & $0.47\pm0.20$ & $29.67\pm0.06$ \\ \hline
    \end{tabular}
    }
    \caption{\textbf{Reconstruction evaluation.} 
    }
    \label{tab:supacc}
\end{table}

\section{Reconstruction examples spanning 32 weeks}

\begin{figure}[!ht]
    \centering\includegraphics[width=0.7\linewidth]{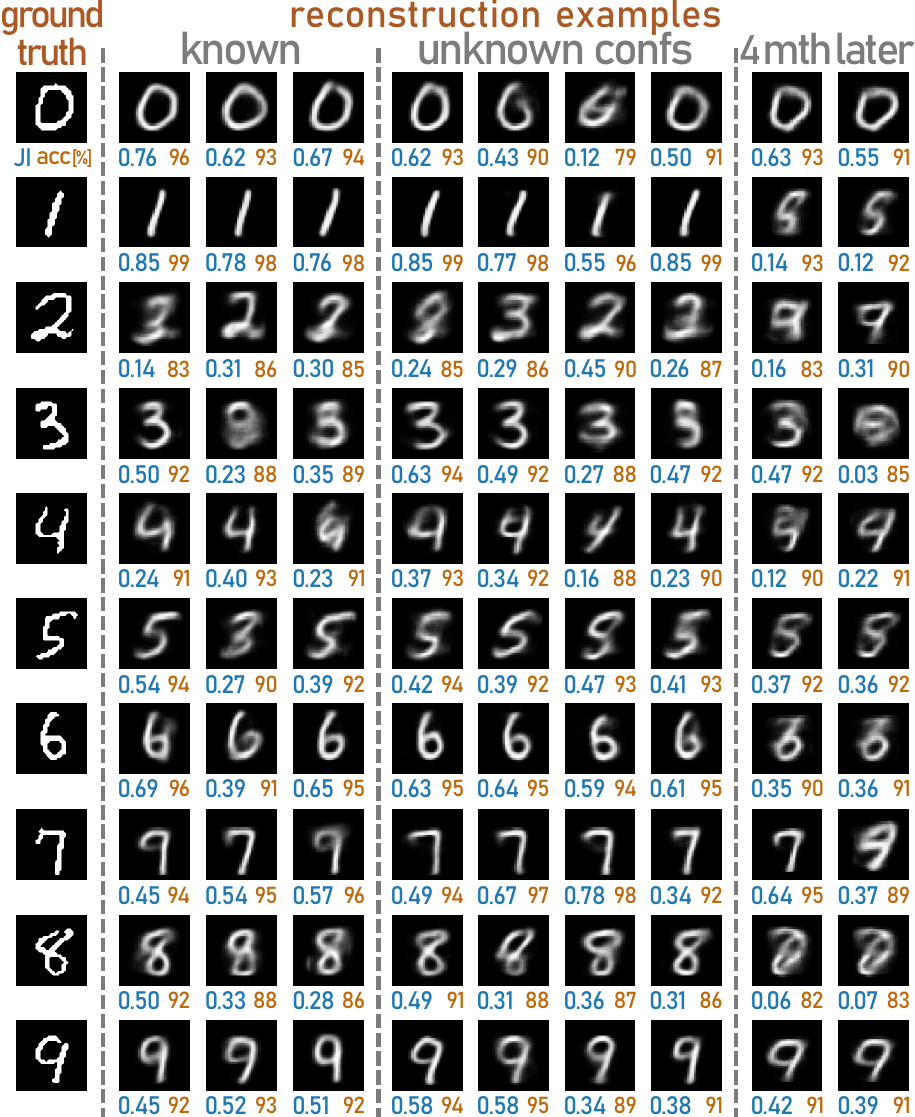}
    \caption{\textbf{Reconstructions produced by configuration training of speckle from multiple low-correlated fiber conformations.}
    Reproduction of Fig. 5 of the main text, accompanied by the reconstructions from two unknown fiber conformations acquired 4 months after the last of the data was acquired for the experiment described in the main text. The trained model was not updated before producing these reconstructions.}
    \label{fig:sup32weeks}
\end{figure}

To evaluate the stability of our configuration-trained model over time, we acquired additional data from fiber configurations 4 months after finishing acquiring the data that was used to train or test our model (as described in the main text). We used the trained model, without additional training or optimization, to create reconstructions from these unknown configurations. In Fig. \ref{fig:sup32weeks} we reproduce Fig. 5 of the main text, with two additional columns that show reconstructions from two low-correlated configurations (right columns). For the most part, results are qualitatively similar to earlier reconstructions and the reconstructions resemble the ground truth, although to a lower degree (in terms of the JI and accuracy) than the configurations acquired before. In total, data was acquired over the span of 32 weeks. This is the first demonstration, to the best of our knowledge, for image reconstruction through a multimode fiber using a single trained model for such a long period of time. Moreover, using configuration training, this is achieved for both known and unknown configurations of a strongly perturbed fiber.

